# Strain-Driven Disproportionation at a Correlated Oxide Metal-Insulator Transition

T. H. Kim,[1] T. R. Paudel,[2] R. J. Green,[3,4] K. Song,[5] H.-S. Lee,[5] S.-Y. Choi,[5] J. Irwin,[6] B. Noesges,[7] L. J. Brillson,[7] M. S. Rzchowski,[6] G. A. Sawatzky,[3] E. Y. Tsymbal,[2] and C. B. Eom[1,*]

[1]*Department of Materials Science and Engineering, University of Wisconsin-Madison, Madison, Wisconsin 53706, USA*

[2]*Department of Physics and Astronomy & Nebraska Center for Materials and Nanoscience, University of Nebraska, Lincoln, Nebraska 68588, USA*

[3]*Quantum Matter Institute, Department of Physics and Astronomy, University of British Columbia, Vancouver, British Columbia, Canada V6T 1Z4*

[4]*Department of Physics & Engineering Physics, University of Saskatchewan, Saskatoon, Saskatchewan, Canada S7N 5E2*

[5]*Materials Modeling & Characterization Department, Korea Institute of Materials Science, Changwon 51508, Republic of Korea*

[6]*Department of Physics, University of Wisconsin-Madison, Madison, Wisconsin 53706, USA*

[7]*Department of Physics, The Ohio State University, Columbus, Ohio 43210, USA*

---

[*]ceom@wisc.edu

**Metal-to-insulator phase transitions in complex oxide thin films are exciting phenomena which may be useful for device applications, but in many cases the physical mechanism responsible for the transition is not fully understood. Here we demonstrate that epitaxial strain generates local disproportionation of the $NiO_6$ octahedra, driven through changes in the oxygen stoichiometry, and that this directly modifies the metal-to-insulator phase transition in epitaxial (001) $NdNiO_3$ thin films. Theoretically, we predict that the Ni-O-Ni bond angle decreases, while octahedral tilt and local disproportionation of the $NiO_6$ octahedra increases resulting in a small band gap in otherwise metallic system. This is driven by an increase in oxygen vacancy concentration in the rare-earth nickelates with increasing in-plane biaxial tensile strain. Experimentally, we find an increase in pseudocubic unit-cell volume and resistivity with increasing biaxial tensile strain, corroborating our theoretical predictions. With electron energy loss spectroscopy and x-ray absorption, we find a reduction of the Ni valence with increasing tensile strain. These results indicate that epitaxial strain modifies the oxygen stoichiometry of rare-earth perovskite thin films and through this mechanism affect the metal-to-insulator phase transition in these compounds.**

Metal-to-insulator phase transitions (MITs) in strongly-correlated electronic systems are fascinating phenomena which have attracted significant attention for decades [1]. Among complex oxide materials which exhibit MITs are rare-earth nickelates having the generic formula $R$NiO$_3$, where the rare-earth element ($R$) is smaller than lanthanum, i.e. $R$ = Pr, Nd ... [2]. The critical temperature of the MIT is dependent on the Ni-O-Ni bond angle: straightening the angle with a larger $R$ cation stabilizes the metallic state over the insulating state and lowers the transition temperature [3–5]. For example, the MIT temperatures in bulk NdNiO$_3$ and SmNiO$_3$ (Ni-O-Ni bond angles of 157.1 and 153.4°, respectively) have been reported to be approximately 200 and 400 K, respectively. It should be also emphasized that a breathing order by disproportionation of the Ni-O bond length plays a crucial role in the MIT [6-10].

In $R$NiO$_3$ thin films, misfit strain arising from a lattice mismatch between the film layer and the underlying substrate affects the lattice volume, electrical conductivity and MIT temperature [11–18]. In particular, films under in-plane tensile strain are more insulating compared to those under in-plane compressive strain. The origin of this interesting phenomenon remains unclear, although a number of mechanisms have been proposed [19-24]. We also note that the effect of oxygen non-stoichiomety on the MIT has been reported in bulk $R$NiO$_3$ [25,26].

It is widely accepted that epitaxial strain in transition metal oxide films can be accommodated through the formation of oxygen vacancy defects, resulting in off-stoichiometry of the compound [27]. Thus, cation/oxygen stoichiometry is an important factor involved with physical and functional properties of complex oxide films [2,28]. A missing cation or oxygen at a given lattice site modifies the local charge/spin/orbital configuration, critically affecting material properties. There have been recent studies on how such vacancy defects inducing off-stoichiometry influence the structural, electronic, magnetic, and transport characteristics of complex oxide films [27–34]. It is therefore possible that the experimentally-observed strain-driven changes in the lattice volume, electrical conductivity and MIT temperature of $R$NiO$_3$ thin films [11–18] are mediated by off-stoichiometry resulting from the formation of oxygen vacancies. Answering this question is of great significance for the understanding of a broader range of phenomena occurring in complex oxide thin films subject to epitaxial strain.

In this work, we present the strain-dependent formation of oxygen vacancies and their impact on the metal-to-insulator phase transition in $R$NiO$_3$ ($R$ = Nd, Sm) (001) thin films. Theory predicts that oxygen vacancies can be created to accommodate biaxial tensile strain in $R$NiO$_3$ films. The oxygen vacancies drive long-ranged structural and electronic modifications, which enhance disproportionation of the NiO$_6$ octahedra stabilizing the insulating phase [35]. Experiment observes that the pseudocubic unit-cell volume and resistivity in epitaxial $R$NiO$_3$ films increase with biaxial tensile strain, which is consistent with our theoretical predictions. It is highly likely that the strain-induced oxygen vacancies produce a more ionic Ni$^{2+}$ state and enhance the local disproportionation. Our results reveal that oxygen stoichiometry susceptible to epitaxial strain can be a key factor to figure out the metal-to-insulator transition in complex oxide thin films.

In order to address the above issue, we performed density-functional theory (DFT) calculations. Our theoretical results show that oxygen stoichiometry in $R$NiO$_3$ is very sensitive to in-plane biaxial strain (Fig. 1(a)). The maximum oxygen vacancy formation energy (calculated using chemical potential corresponding to O$_2$ molecule/$R$NiO$_3$ in the dilute limit) decreases as the in-plane strain increases (Supplemental Fig. S1) [36]. For in-plane tensile strain above 1.9 % (2.2 %), the formation energy of an oxygen vacancy in NdNiO$_3$ (SmNiO$_3$) becomes negative and therefore, the materials become largely oxygen deficient at large tensile strain (Fig. 1a), while they are stoichiometric for compressive strain.

Further, we find that an oxygen vacancy is more easily formed in the NiO$_2$ plane than in the RO (R = Nd, Sm) plane of the RNiO$_3$ unit cell [36]. When the oxygen atom connecting two octahedra is removed in the NiO$_2$ plane, both in-plane octahedra tilt towards each other (Fig. 1(b)). The oxygen octahedral structure of bulk NdNiO$_3$, is represented by the tilt angle $\theta$ with respect to the $z$ axis and rotation angle $\phi$ in the $xy$ in-plane (the inset of Fig. 1(b)). As in-plane biaxial strain increases, $\theta$ clearly decreases while $\phi$ remains nearly constant, indicating the predominant effect of strain on octahedral tilting [37]. In the presence of oxygen vacancies, NiO$_6$ octahedra in $R$NiO$_3$ are more tilted, as reflected in smaller $\theta$ values, whereas $\phi$ is approximately unchanged.

In addition to octahedral rotations and tilts, we also observe disproportionation of the NiO$_6$ octahedra. There are two inequivalent octahedra (inset in Fig. 1(b)) with shorter and longer Ni-O bond lengths, as shown in Fig. 1(c). For stoichiometric NdNiO$_3$ (Fig. 1(c) top panel), our DFT calculations indicate that this disproportionation decreases with the increasing in-plane biaxial strain and eventually vanishes making the two octahedra equivalent beyond epitaxial strain of 1 %, at which two different Ni-O bonds in a-b plane and c plane at compressive strain merge to single a-b plane and c-plane Ni-O bonds. In the presence of oxygen vacancies (Fig. 1(c) lower panel), the disproportionation is more pronounced, and persists for all investigated levels of strain.

This change in disproportionation with oxygen vacancy dramatically affects the overall electronic properties. The cyan background in Fig. 1(d) shows the zero strain, stoichiometric, total density of states (DOS). It indicates an overall metallicity, with near band-edge states that are dominated by mixed O-2$p$ and Ni-3$d$ bands. The solid red line in the same panel shows the opening of a band gap when oxygen vacancies are included, driving a metal-insulator transition. The DOS (the dark yellow line) shows that occupied defect states lie in the valence band of the host material and, hence, the defect-induced effects are long-ranged. This is in contrast to wide band gap semiconductors where defect states lie in the band gap and their effect is short-ranged. The long-ranged effect of oxygen vacancies promotes the octahedral tilting and produces additional disproportionation and drives the system into the insulating state. The metallicity of the defect-free system suggests that the DFT does not completely capture the electronic properties, but the calculation clearly indicates that vacancy-induced disproportionation drives the system toward the insulating state, which would result in a higher MIT temperature. For SmNiO$_3$, the DFT calculations show qualitatively analogous behavior.

To study the effect of in-plane biaxial strain on oxygen stoichiometry, lattice structures, and electronic properties, we grew epitaxial $R$NiO$_3$ ($R$ = Nd, Sm) (001) films using pulsed laser deposition (PLD) with *in situ* monitoring by reflection high energy electron diffraction (RHEED). During the PLD growth of $R$NiO$_3$ (001) films, RHEED oscillations were used to estimate the film thickness (Supplemental Fig. S3) [36]. In Fig. 2(a), atomic force microscopy (AFM) topography and RHEED pattern images show that the as-grown NdNiO$_3$ (14 nm) films have atomically-flat surfaces with a clear step-terrace structure and high crystalline qualities, respectively. As depicted in Fig. 2(b), LaAlO$_3$ (LAO) (001), NdGaO$_3$ (NGO) (110), and SrTiO$_3$ (STO) (001) substrates were used to investigate epitaxial films in a wide range of strain states. Note that the pseudocubic lattice constants of bulk NdNiO$_3$ and SmNiO$_3$ are 3.808 and 3.801 Å, respectively [5]. Thus, the lattice mismatch of an NdNiO$_3$ (SmNiO$_3$) film is -0.5 (-0.3), 1.3 (1.5), and 2.5 (2.7) % with respect to pseudocubic LAO, NGO, and STO substrates, respectively, that is, under in-plane compressive strain for the LAO substrate and under in-plane tensile strain for the NGO and STO substrates. To identify the coherency and actual strain states, reciprocal space mappings (RSMs) around the pseudocubic (-103) Bragg peaks were performed, as shown in Fig. 2(c) (For SmNiO$_3$ (001) films, see Supplemental Fig. S4(a).) [36]. These show that the films are coherent with respect to the underlying substrate and strained with the same in-plane lattice constant as that of the substrate in pseudocubic notation. By deriving a pseudocubic unit-cell volume from the measured lattice constants, we also examined the relation between structural properties and oxygen off-stoichiometry in $R$NiO$_3$ films.

As evident from Fig. 2(d), the pseudocubic unit-cell volume of $R$NiO$_3$ films increases with in-plane tensile strain, whereas it is identical to the bulk value for in-plane compressive strain (Supplemental Tab. S1) [36]. We found that the measured out-of-plane lattice constants deviate from those calculated assuming the volume conservation of $R$NiO$_3$ unit cells. This difference between the measured and calculated lattice constants gets larger with in-plane tensile strain, indicative of a volume expansion (Supplemental Fig. S5) [36]. The volume expansion of $R$NiO$_3$ unit cells by in-plane tensile strain is also confirmed in atomic-scale scanning transmission electron microscopy (STEM) measurements (Supplemental Fig. S6) [36]. It is further interesting that an NdNiO$_3$ film on a STO substrate (a lattice mismatch of +2.5 %) is coherent with a volume expansion, whereas an NdNiO$_3$ film on a YAlO$_3$ substrate (a lattice mismatch of -2.7 %) is relaxed, preserving the unit-cell volume (Supplemental Fig. S7) [36]. For NdNiO$_3$ and SmNiO$_3$ films grown on STO (001) substrates, their unit-cell volumes increase up to 4.4 and 6.1 %, respectively. *Albeit* the strain dependence of the Poisson's ratio is considered, this volume change in $R$NiO$_3$ films is too large. Note that the Poisson's ratio in complex oxides is usually in the range of 0.2~0.3 [38,39]. This is suggestive of a change in oxidation state under tensile strain, but not compressive strain. Similar volume expansion by in-plane tensile strain has been previously reported in other transition-metal oxide films where multiple oxidation states of the transition-metal element are allowed [40,41].

We also found that electronic transport properties strongly depend on epitaxial strain. Figure 2(e) shows the temperature dependence of resistivity in the 14-nm-thick $NdNiO_3$ (001) films. $NdNiO_3$ samples undergo an explicit MIT as the temperature decreases. It is evident that $NdNiO_3$ films under in-plane tensile strain (grown on NGO and STO substrates) exhibit larger resistivity than those under in-plane compressive strain (grown on a LAO substrate) at all temperatures. In $SmNiO_3$ films, similar transport behaviors are observed (Supplemental Fig. S4(b)) [36]. In particular, the MIT in the tensile-strained $NdNiO_3$ films occurs at a higher temperature, whereas it is suppressed in the compressive-strained $NdNiO_3$ films with a lower MIT temperature (Supplemental Fig. S8) [36]. And, the $NdNiO_3$/LAO film with better electrical conductivity gave higher carrier concentrations of mobile charges than the $NdNiO_3$/STO film (Supplemental Fig. S9) [36]. These strain-dependent transport properties are consistent with previous reports [11–14,17,18].

Strain-induced oxygen vacancies increase a pseudocubic unit-cell volume and resistivity in $R$NiO$_3$ films. By carrying out cathodoluminescence (CL) measurements of our as-grown $NdNiO_3$ (001) films (Supplemental Fig. S10) [36], we first found that the formation of oxygen vacancies in $R$NiO$_3$ films is quite dependent on the in-plane biaxial strain. In the tensile-strained $NdNiO_3$ films (i.e. $NdNiO_3$/$NdGaO_3$ and $NdNiO_3$/$SrTiO_3$ films) unlike the compressive-strained $NdNiO_3$/$LaAlO_3$ film, we only observed distinct CL signals at the low photon energies of 1.8 eV which are directly related with oxygen-vacancy-mediated optical transitions [42,43]. By visualizing oxygen atoms in annular bright field (ABF)-STEM images of $NdNiO_3$/STO and $NdNiO_3$/LAO films, we also verified that the tensile-strained $NdNiO_3$ film is more oxygen-deficient than the compressive-strained $NdNiO_3$ film (Supplemental Fig. S11) [36]. Then, by performing *in situ* annealing experiments of as-deposited $NdNiO_3$ films in different oxygen-ambient environments, it was further identified that more oxygen-deficient $NdNiO_3$ films exhibit a larger unit-cell volume in pseudocubic notation and higher resistivity (Supplemental Fig. S12) [36]. A possibility of cation vacancies in this volume expansion could be excluded by examining the cation stoichiometry of as-grown $NdNiO_3$ films using Rutherford backscattering spectrometry (RBS) (Supplemental Fig. S13) [36]. For two $NdNiO_3$ films, one with in-plane compressive strain and the other with in-plane tensile strain, the ratio of chemical composition between Nd and Ni atoms is 1:1 within measurement errors. It is highly likely that the oxygen vacancy concentration in tensile-strained $R NiO_3$ films is higher than that in compressive-strained $R NiO_3$ films. Then, the more amount of oxygen vacancy defects in the tensile-strained $R NiO_3$ films leads to further enlarged volume and further enhanced resistivity.

We observe a reduction of Ni valence state in 10-unit-cell-thick $NdNiO_3$ (3.8 nm) films under in-plane tensile strain using the atomic-resolved STEM and electron energy loss spectroscopy (EELS) techniques. For two epitaxial $NdNiO_3$ thin films with atomically sharp interfaces grown on STO (2.5% in-plane tensile strain) and LAO (0.5 % in-plane compressive strain) (001) substrates, as shown in Figs. 3(a) and 3(b), the Ni $L_{2,3}$-edges were measured over the film thickness (Figs. 3(c) and 3(d)). All EELS spectra in each $NdNiO_3$ (001) film are similar with respect to the thickness, which indicates that the electronic and chemical states of Ni in the as-grown $NdNiO_3$ film are spatially homogeneous throughout the whole thickness. On the other hand, the comparison of Ni $L_2$-edges between NdNiO$_3$/STO and NdNiO$_3$/LAO films (Fig. 3(e)) makes sure that the valence states of Ni cations are strongly correlated with the in-plane biaxial strain. Further confirmation with EELS calculation of Ni$^{2+}$ and Ni$^{3+}$ valence states reveals that the oxidation states of Ni cations in the NdNiO$_3$/STO and NdNiO$_3$/LAO films are more towards ionic Ni$^{2+}$ and covalent Ni$^{3+}$, respectively.

Figure 4 shows Ni $L_{2,3}$ x-ray absorption spectroscopy (XAS) data for the 10-unit-cell-thick NdNiO$_3$ (3.8 nm) films on different substrates. For each case, spectra are measured at 22 K, where the nickelate films are insulating. In the insulating phase, a two-peaked structure is present at the Ni $L_3$ edge (between ~853 and ~854 eV). It is interesting that the XAS spectra of the NdNiO$_3$ films become closer to that of NiO, as the substrate changes from LAO to NGO to STO. While the first sharp peak (denoted by the dotted line in Fig. 4) is an intrinsic feature of the nickelate spectrum [6,44–46], the identical and sharp $L_3$ peak of Ni$^{2+}$ in NiO shifts negligibly lower in energy. In thicker (14 nm) nickelate films (Supplemental Fig. S14(a)) [36], though there is slightly less Ni$^{2+}$ observed via XAS, a similar increase in the first peak intensity is observed for increasing tensile strain. This indicates that the Ni valence is reduced toward Ni$^{2+}$ for tensile strain. In an oxygen-deficient NdNiO$_3$ film, the reduction of the Ni valence state is more pronounced (Supplemental Fig. S15) [36]. Oxygen vacancies formed to accommodate in-plane tensile strain reduce the oxidation state of Ni cations from the covalent Ni$^{3+}$ to the ionic Ni$^{2+}$ in NdNiO$_3$ (001) films, which promotes disproportionation of Ni 3$d$-O 2$p$ hybridization (Supplemental Figs. S14(b) and S14(c)) [36].

In conclusion, we have demonstrated that strain-driven changes in structural distortion, lattice volume, electrical conductivity and a MIT temperature of epitaxially grown rare-earth nickelate films are mediated by chemical off-stoichiometry resulting from the formation of oxygen vacancies. This result is of significance for fundamental understanding of a broader range of physical properties and phenomena occurring in complex oxide thin films subject to epitaxial strain.

This work was supported by the National Science Foundation (NSF) under DMREF Grant No. DMR-1629270. The research at University of Nebraska-Lincoln was supported by the NSF through the Nebraska Materials Research Science and Engineering Center under Grant No. DMR-1420645. The work at University of British Columbia was supported by the Natural Sciences and Engineering Research Council of Canada (NSERC) and Canada Research Chairs program. Parts of the research were performed at the Canadian Light Source, which is funded by the Canada Foundation for Innovation, NSERC, the National Research Council Canada, the Canadian Institutes of Health Research, the Government of Saskatchewan, Western Economic Diversification Canada, and the University of Saskatchewan. The research was also supported by the Global Frontier Program through the Global Frontier Hybrid Interface Materials (GFHIM) of the National Research Foundation of Korea funded by the Ministry of Science, ICT & Future Planning (2013M3A6B1078872).

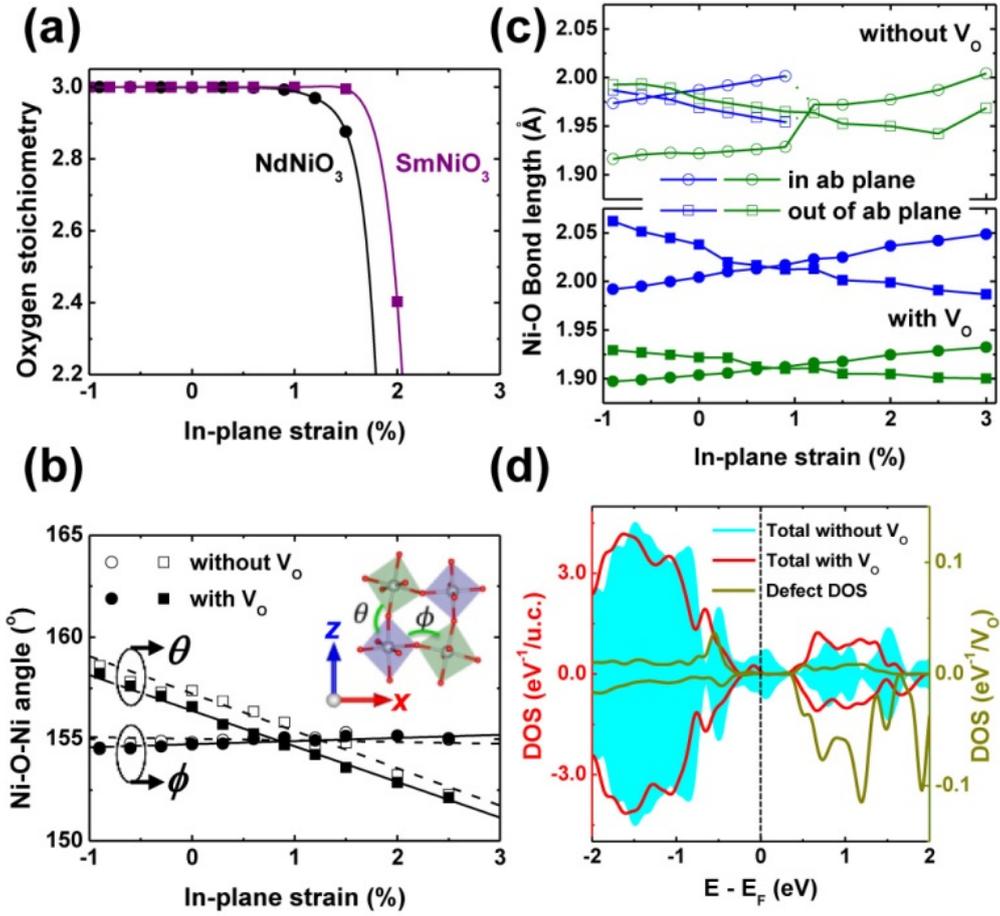

FIG. 1 (color online). (a) Strain-dependent oxygen stoichiometry in NdNiO$_3$ (solid circles) and SmNiO$_3$ (solid squares). The stoichiometry is calculated at room temperature and thermal equilibrium with O$_2$ molecule. The lines are guides to the eye for the data points. (b) Calculated octahedral tilt ($\theta$) and rotation ($\phi$) angles as a function of in-plane strain. The inset shows a *Pbnm* NdNiO$_3$ unit cell indicating the angles $\theta$ and $\phi$. The tilt angle $\theta$ and rotation angle $\phi$ represent the Ni-O-Ni bonding angles with respect to the *z* axis and in the *xy* in-plane, respectively. In (b), open (filled) symbols represent the quantities in the absence (presence) of oxygen vacancies. (c) Calculated shorter (green) and longer (blue) Ni-O bond lengths in a-b plane (circle) and c plane (square) in absence (open symbol) and presence (filled symbol) of oxygen vacancy. Note that the difference between the shorter and longer Ni-O bondings increases in presence of oxygen vacncy. (d) Total and local density of states (DOS) in NdNiO$_3$ projected at the defect sites in the presence (lines) and absence (cyan background) of oxygen vacancies. Note a band gap for the oxygen-deficient NdNiO$_3$.

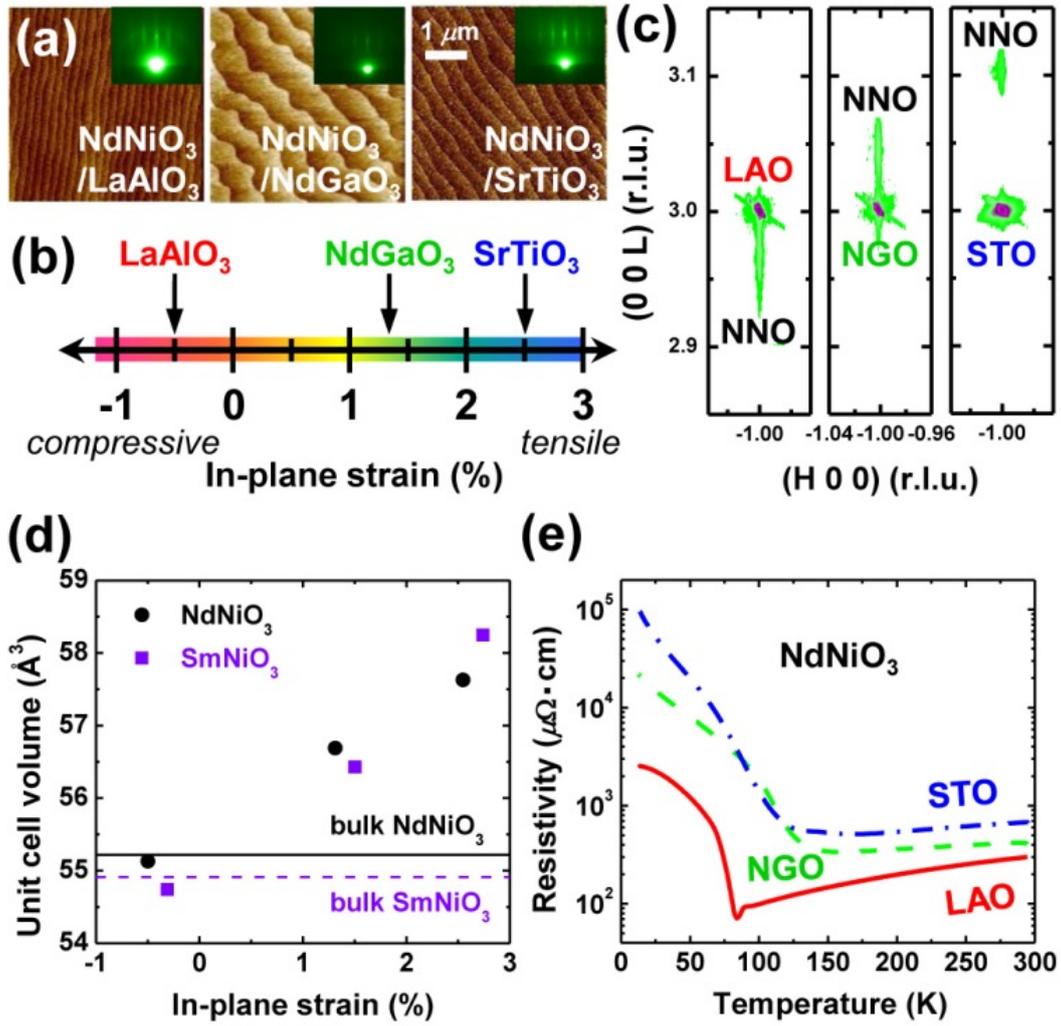

FIG. 2 (color online). (a) AFM topography images of the as-grown 14-nm-thick (left) NdNiO$_3$/LaAlO$_3$ (LAO), (middle) NdNiO$_3$/NdGaO$_3$ (NGO), and (right) NdNiO$_3$/SrTiO$_3$ (STO) thin films, where the insets represent the corresponding RHEED patterns. (b) In-plane biaxial strain of NdNiO$_3$ with respect to pseudocubic LAO, NGO, and STO (001) substrates. (c) (-103) RSMs of pseudocubic NdNiO$_3$ (001) films. All of the RSMs are on the same scale in the *H* and *L* directions. All NdNiO$_3$ films are coherent with respect to the substrates. (d) The strain-dependent pseudocubic unit-cell volumes of epitaxial NdNiO$_3$ (solid circles) and SmNiO$_3$ (solid squares) films. The solid and dashed lines represent the unit-cell volume of bulk NdNiO$_3$ and SmNiO$_3$ in pseudocubic notation, respectively. (e) The temperature-dependent resistivity in the 14-nm-thick NdNiO$_3$ (001) films.

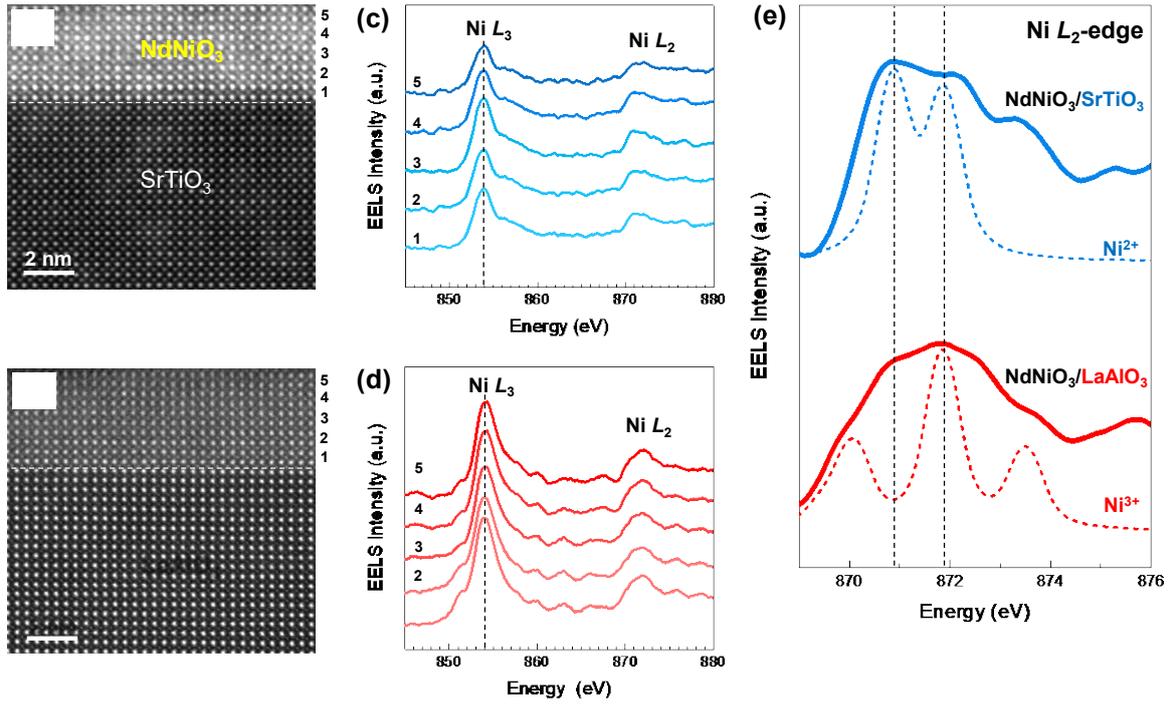

FIG. 3 (color online). STEM-high angle annular dark field (HAADF) images of the 10-unit-cell-thick $NdNiO_3$ (3.8 nm) films grown on (a) STO and (b) LAO substrates along the pseudocubic [100] zone axis. (c,d) EELS spectra of Ni $L_{2,3}$-edges obtained across the $NdNiO_3$ films shown in (a) and (b), respectively. Black dashed and gray dotted lines are guidelines for Ni $L_3$. (e) The Ni $L_2$-edges of the $NdNiO_3$/STO (the blue solid line) and $NdNiO_3$/LAO (the red solid line) films in (c) and (d), respectively. A comparison with the EELS calculation spectra of $Ni^{2+}$ (blue dashed line) and $Ni^{3+}$ (red dashed line) evidences that the oxidation states of Ni in 10-unit-cell-thick $NdNiO_3$ films grown on STO and LAO substrates are more towards $Ni^{2+}$ and $Ni^{3+}$, respectively. For $Ni^{2+}$, two distinct peaks (870.8 and 871.9 eV) was calculated with the estimated crystal field splitting parameter, 10 Dq, 1.1 eV, while three peaks (870.1, 871.9, and 873.8 eV) with higher 10 $D_q$, 1.8 eV was found in the $Ni^{3+}$ state.

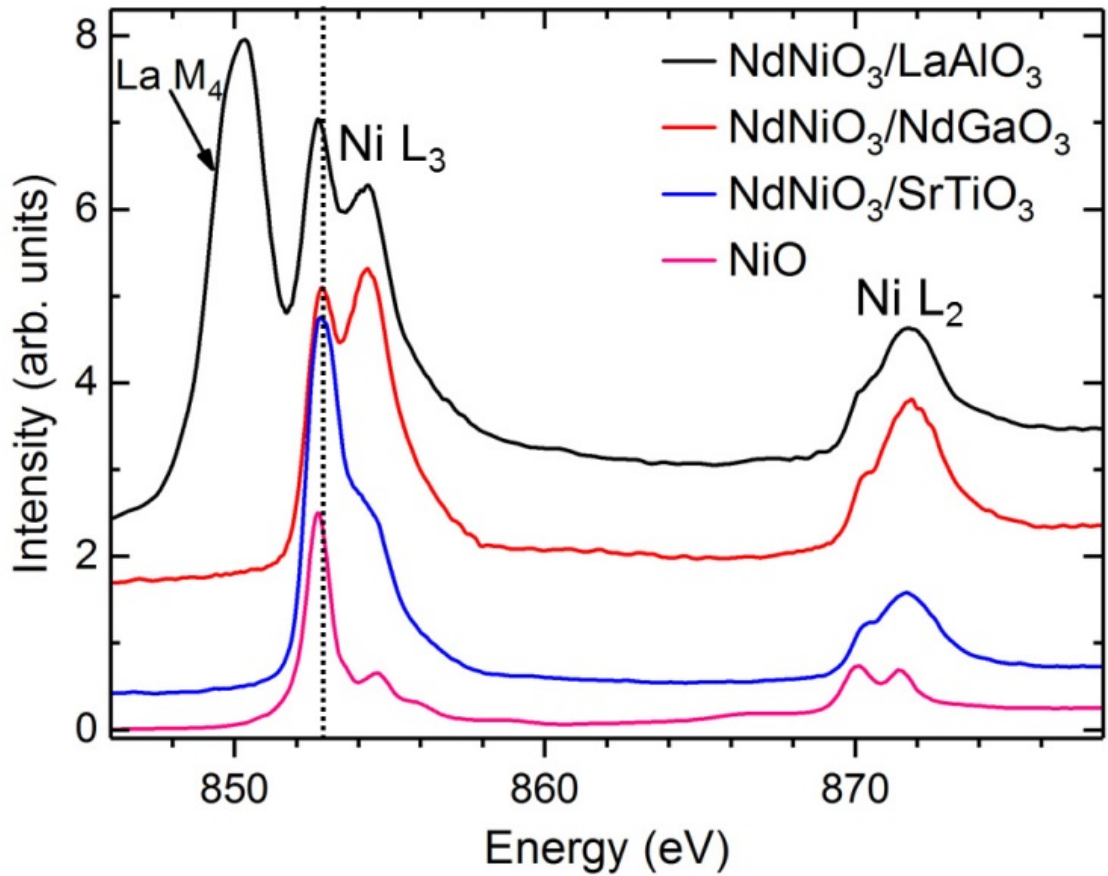

FIG. 4 (color online). Ni $L_{2,3}$ x-ray absorption spectroscopy (XAS) of the 10-unit-cell-thick NdNiO$_3$ (3.8 nm) films on LAO, NGO, and STO substrates at 22 K. All spectra are measured via the total fluorescence yield (TFY) mode, which is bulk-sensitive with a probing depth of several tens of nanometers. The XAS spectra of the NdNiO$_3$ films become closer to that of NiO, as the substrate changes from LAO to NGO to STO, indicative of the reduction of Ni charge valence toward ionic Ni$^{2+}$ by in-plane tensile strain.

**Method of theoretical calculations**

The formation energy of oxygen vacancy is calculated using, $\Delta H_f(V_O) = E(d) - E_H + \mu_O$, where $E(d)$ and $E(H)$ are the energy of the system with and without the defect, and $\mu_O$ is the oxygen chemical potential defined with respect to its elemental form ($1/2E(O_2)$) in the triplet state. We include an *ad hoc* correction energy of +0.233 eV [1,2] to the calculated elemental energy to accurately describe the thermodynamic properties. We use a 160-atom 2×2×2 supercell of orthorhombic rare-earth nickelates, sufficiently large to avoid inter-defect interaction. The cell has four Ni-O planes each containing eight octahedra. We remove one oxygen from the supercell to simulate the defect concentration in the dilute limit. The out-of-plane lattice parameter and internal atomic coordinates of the defective cell are relaxed until the Hellman Feynman force on each atom is less than 0.02eV/Å to determine the associated bond angle (Fig 1b), the bond length (Fig 1c) and the defect formation energy for each biaxial strain defined with respect to the experimental lattice constant *a*. The oxygen vacancy concentration is the calculated using, $N(V_O) = N_A \exp(-\Delta H/K_B T)$ where $N_A$ is the total number of oxygen sites. The calculations are performed using density functional theory (DFT) implemented within Vienna *ab initio* simulation package (VASP) [3,4] utilizing the projected augmented wave (PAW) method [5] and Perdew-Burke-Ernzerhof (PBE) pseudopotentials [6]. We fully relax the atomic structure with force convergence limit of 0.01eV/atom. Correlation effects beyond the Generalized Gradient Approximation (GGA) are treated at semi-empirical GGA+*U* level [7] in a rotationally invariant formalism [8] with $U$ = 3.0 eV for the Ni-3*d* orbitals. The *f*-electrons of the rare-earth elements are considered as part of core electrons. The $V_O$ resolved density of states in Fig. 1(d) is calculated by projecting the total density of states into the sphere of radius 1 Å centered at the defect site.

**Method of scanning transmission electron microscopy (STEM)-electron energy loss spectroscopy (EELS) measurements**

Cross-sectional TEM samples for STEM-EELS were prepared by mechanical grinding to a thickness and dimpling to a thickness. The dimpled samples were then ion-milled at liquid nitrogen temperature using a first a 4 kV $Ar^+$ ion beam (PIPS, Gatan, Inc) and then a low energy $Ar^+$ ion beam to remove surface damage. To reduce electron beam damage, high angle annular dark field (HAADF) STEM images were acquired at 120 kV accelerating voltage in a STEM (JEM-2100F, JEOL) equipped with a spherical aberration corrector (CEOS GmbH). The probe convergence angle of approximately 22 mrad was used. The inner and outer angles of the HAADF detectors were 90 and 200 mrad, respectively. The Wiener filter was used to reduce background noise from STEM image (HREM Research Inc., Japan). EEL spectra were also obtained at 120 kV using an EEL spectrometer (Gatan GIF Quantum ER, USA) with an energy resolution of 0.8 eV.

**Method of EELS calculations**

A relativistic configuration-interaction (CI) calculations of the Ni-$L_{2,3}$ edge in $NdNiO_3$ (*Pbmn*, $a$ = 5.38 Å, $b$ = 5.37Å, $c$ = 7.60 Å) were performed. The model clusters, $(NiO_6)^{9-}$ for $Ni^{3+}$ and $(NiO_6)^{10-}$ for $Ni^{2+}$ were constructed based on the crystal structure. The relativistic molecular orbital (MO) calculations were calculated for each cluster to find the ground state and excited state configurations. For $Ni^{3+}$, the transition from $(2p_{1/2})^2(2p_{3/2})^4(t_{2g})^6(e_g)^1$ to $(2p_{1/2})^2(2p_{3/2})^3(t_{2g})^6(e_g)^2$ in many-electron picture was considered while $(2p_{1/2})^2(2p_{3/2})^4(t_{2g})^6(e_g)^2$ to $(2p_{1/2})^2(2p_{3/2})^3(t_{2g})^6(e_g)^3$ for $Ni^{2+}$. The oscillator strength of electric-dipole transition over all polarization was calculated between ground state and excited state. The final spectrum was constructed with 0.8 eV Gaussian broadening. According to the configuration analysis of the many-electron wave functions, the major peaks mainly corresponds to $2p_{1/2} \rightarrow e_g$ transition. The calculated $Ni^{2+}$ $L_2$-edge EELS in this study is reasonably consistent with the previous study with NiO [9]. Although more careful consideration and comparison are necessary to confirm the intensity and peak positions, it is worth to mention that triple peaks for $Ni^{3+}$ and double peaks for $Ni^{2+}$ was calculated in NiO nanocrystalline [10].

**Supplemental TAB. S1.** In-plane misfit strain, in-plane (a), out-of-plane (c) lattice constants, and unit cell volumes of epitaxial $R$NiO$_3$ (001) ($R$ = Nd, Sm) films and bulk $R$NiO$_3$ [Ref. 5 in the manuscript].

| film/subs. | strain (%) | $a$ (Å) | $c$ (Å) | u.c. volume (Å$^3$) |
|---|---|---|---|---|
| NdNiO$_3$/LaAlO$_3$ | -0.5 | 3.789 | 3.840 | 55.13 |
| NdNiO$_3$/NdGaO$_3$ | +1.3 | 3.858 | 3.809 | 56.69 |
| NdNiO$_3$/SrTiO$_3$ | +2.5 | 3.905 | 3.779 | 57.63 |
| Bulk NdNiO$_3$ | 0.0 | 3.808 | 3.808 | 55.22 |
| SmNiO$_3$/LaAlO$_3$ | -0.3 | 3.789 | 3.813 | 54.74 |
| SmNiO$_3$/NdGaO$_3$ | +1.5 | 3.858 | 3.791 | 56.43 |
| SmNiO$_3$/SrTiO$_3$ | +2.7 | 3.905 | 3.820 | 58.25 |
| Bulk SmNiO$_3$ | 0.0 | 3.801 | 3.801 | 54.91 |

**Strain dependence of oxygen vacancy formation energy in rare-earth nickelate (*R*NiO₃) films**

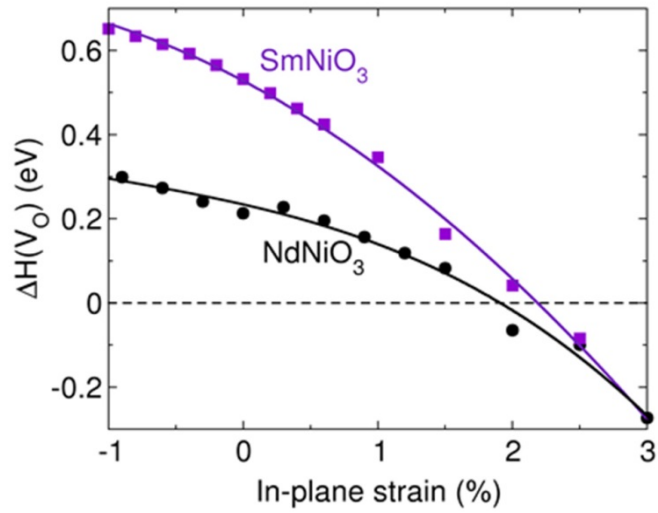

**Supplemental FIG. S1** (color online). Oxygen vacancy ($V_O$) formation energy as a funtion of in-plane strain measured with respect to bulk experimental lattice constants in NdNiO₃ (solid circles) and SmNiO₃ (solid squares). The oxygen vacancy formation energy is calculated at room temperature and thermal equilibrium with O₂ molecule. The filled circle and square represents data points and the lines are guide to an eye.

**Oxygen-vacancy-mediated strain stabilization in rare-earth nickelate ($R$NiO$_3$) films**

In $R$NiO$_3$ thin-film heterostructures, two different processes for epitaxial stabilization are expected and the related physical properties should be different depending on in-plane biaxial strain. Under in-plane compressive strain, a unit cell elongates along the out-of-plane direction (*i.e.* [001]) with changes of both out-of-plane and in-plane lattice constants, as shown in Supplemental Fig. S2(a). The chemical composition should be identical to the bulk composition without the formation of vacancy defects. Then, the volume of unit cells is preserved to the bulk value. In contrast, for in-plane tensile strain (Supplemental Fig. S2(b)), oxygen vacancies are spontaneously created to accommodate the biaxial strain from the underlying pseudocubic substrates resulting in oxygen off-stoichiometry. Note that $R$NiO$_3$ is a layered compound containing alternating layers of $R$O ($R$ = Nd, Sm) and NiO$_2$ along the [001] direction. Considering that the in-plane tensile strain stretches a Ni-O-Ni bond horizontally, an oxygen vacancy is expected to form easier in a NiO$_2$ plane than a $R$O plane. When an oxygen vacancy is created at the site connecting two oxygen octahedra in the NiO$_2$ plane, both oxygen octahedra tilt toward each other due to the Coulomb repulsion between the neighboring Ni ions. With changes in the Ni-O bonding length, the unit-cell volume expands beyond the Poisson's ratio. The tilting of NiO$_6$ octahedra may narrow Ni 3$d$ bands further and thereby, the charge transfer gap between Ni 3$d$ and O 2$p$ bands can be opened. It follows that an insulating phase is stabilized in the $R$NiO$_3$ thin films under in-plane tensile strain.

In addition, in unstrained NdNiO3, we find that the calculated formation energy of an oxygen vacancy in the NiO$_2$ plane is lower than in the NdO plane by about 1.7eV. This can be understood in terms of defect induced relaxation and extent of charge disproportionation. In the NdO plane, oxygen is surrounded by the two Ni atoms along the out-of-plane direction, which are free to move when the defect is introduced. On the contrary, in the NiO$_2$ plane, oxygen is bonded to the two in-plane Ni atoms, and when the defect is created their relaxation is arrested by biaxial strain. As the result the Ni atoms undergo more disproportion leading to a larger reduction in energy.

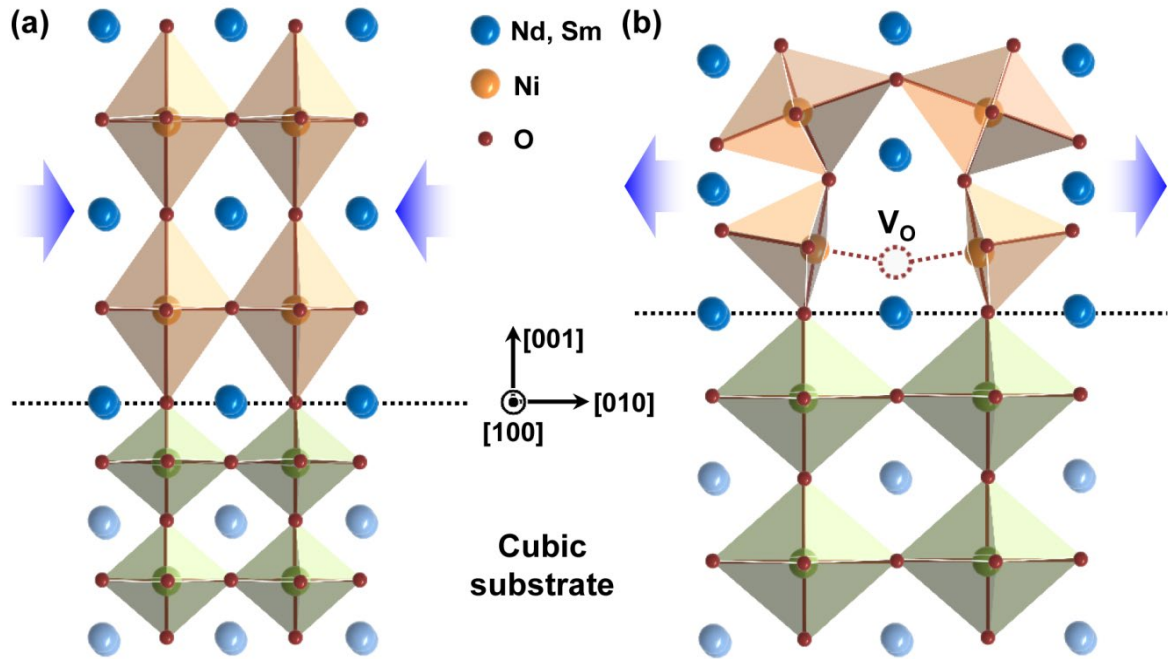

**Supplemental FIG. S2** (color online). Schematic diagrams of strain accommodation mechanisms in $R$NiO$_3$ films under in-plane (a) compressive and (b) tensile strain. For in-plane compressive strain, the unit cells elongate along the out-of-plane direction with volume conservation. In contrast, for in-plane tensile strain, oxygen vacancies (V$_O$) are formed to stabilize epitaxial strain with volume expansion.

**Epitaxial growth of NdNiO$_3$ (001) films**

Epitaxial NdNiO$_3$ thin films on LaAlO$_3$, NdGaO$_3$, and SrTiO$_3$ substrates were grown through pulsed laser deposition (PLD) with *in-situ* monitoring by reflection high energy electron diffraction (RHEED). The growth temperature and oxygen partial pressure were around 550 °C and 0.15 mbar, respectively. In a cooling process after the film deposition, *in-situ* post-annealing was performed under the oxygen atmosphere pressure of 1 atm. During the PLD growth of the NdNiO$_3$ (001) films, a clear RHEED oscillation was observed allowing us to estimate the film thickness (Supplemental Fig. S3), which was confirmed through x-ray reflection (XRR) measurements as well.

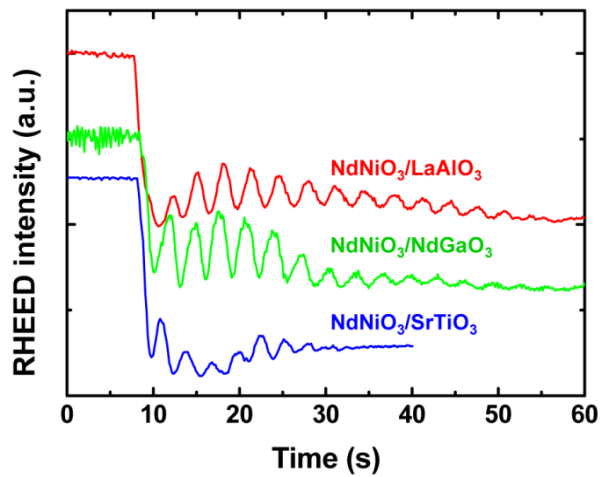

**Supplemental FIG. S3** (color online). The RHEED oscillation of NdNiO$_3$ (001) films during the PLD film growth.

# Epitaxial growth of SmNiO₃ (001) films

In addition to NdNiO$_3$ films, we epitaxially grew SmNiO$_3$ (001) films on LaAlO$_3$, NdGaO$_3$, and SrTiO$_3$ substrates using PLD. To check the coherency of SmNiO$_3$ films, we performed reciprocal space mappings (RSMs) around the pseudocubic (-103) Bragg peaks (Supplemental Fig. S4(a)). The transport properties were also measured in the *Van der Pauw* geometry (Supplemental Fig. S4(b)).

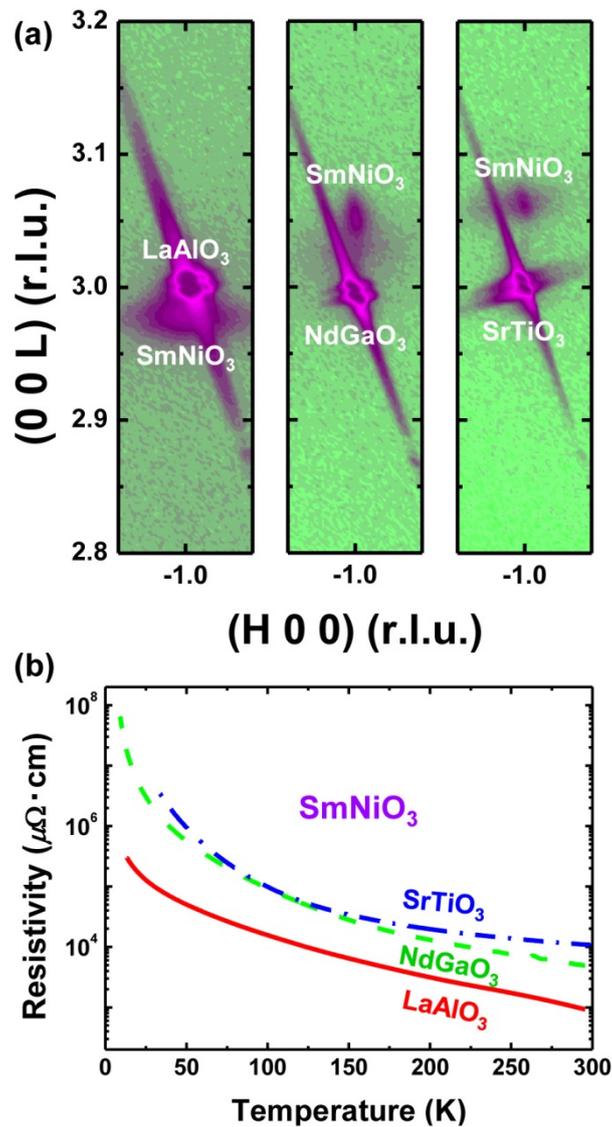

**Supplemental FIG. S4** (color online). (a) (-103) RSMs of pseudocubic SmNiO$_3$ (001) films grown on LaAlO$_3$ (left), NdGaO$_3$ (middle), SrTiO$_3$ (right) substrates. (b) Temperature-dependent resistivity ($\rho$) in epitaxial SmNiO$_3$ (001) films.

**Strain dependence of out-of-lattice constants in NdNiO$_3$ and SmNiO$_3$ (001) films**

In order to calculate unit-cell volumes of NdNiO$_3$ and SmNiO$_3$ films, we measured their out-of-plane lattice constants using x-ray diffraction (XRD) and RSM techniques. We found that the measured out-of-plane lattice constants deviate from those (the solid and dashed lines) calculated assuming the volume conservation of $R$NiO$_3$ unit cells, as in-plane strain increases (Supplemental Fig. S5).

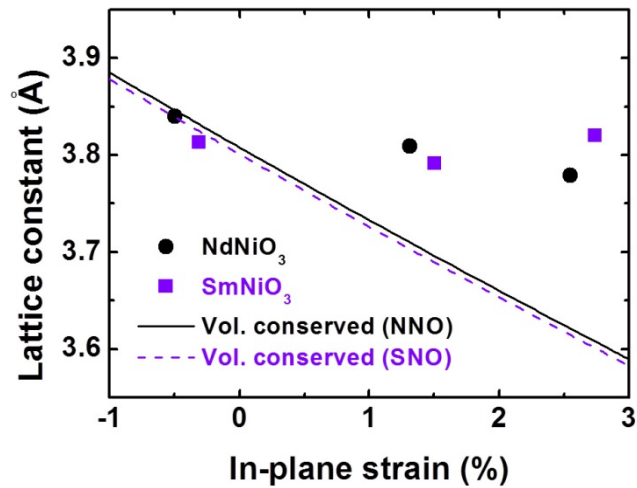

**Supplemental FIG. S5** (color online). Out-of-plane lattice constants of NdNiO$_3$ (solid circles) and SmNiO$_3$ (solid squares) films coherently grown on substrates. The out-of-plane lattice constants of NdNiO$_3$ (the solid line) and SmNiO$_3$ (the dashed line) calculated assuming the volume conservation are plotted as a function of in-plane strain.

**Atomic-level verification of NdNiO$_3$ (001) films under in-plane tensile strain**

To confirm a unit-cell-volume expansion of $R$NiO$_3$ films under in-plane tensile strain in an atomic scale (Supplemental Figs. S6(a) and S6(b)), we extracted thickness-dependent in-plane and out-of-plane lattice constants in scanning transmission electron microscopy (STEM) images of NdNiO$_3$ (001) films grown on LaAlO$_3$ (in-plane compressive strain) and SrTiO$_3$ (in-plane tensile strain) substrates (Supplemental Figs. S6(c) and S6(d)). Then, we measured the volumes of the NdNiO$_3$ unit cell in both NdNiO$_3$/LaAlO$_3$ and NdNiO$_3$/SrTiO$_3$ films as a function of the film thickness (Supplemental Fig. S6(e)). The measured unit-cell volume expands in the NdNiO$_3$/SrTiO$_3$ film, whereas it is close to the bulk value in the NdNiO$_3$/LaAlO$_3$ film. This strain-dependent volume expansion in an atomic level is quite consistent with our structural results obtained by XRD $\theta$-$2\theta$ scans and RSM measurements macroscopically. The oxygen vacancies are also quantitatively verified by Annular Bright Field (ABF) STEM imaging as shown in Supplemental Fig. S12.

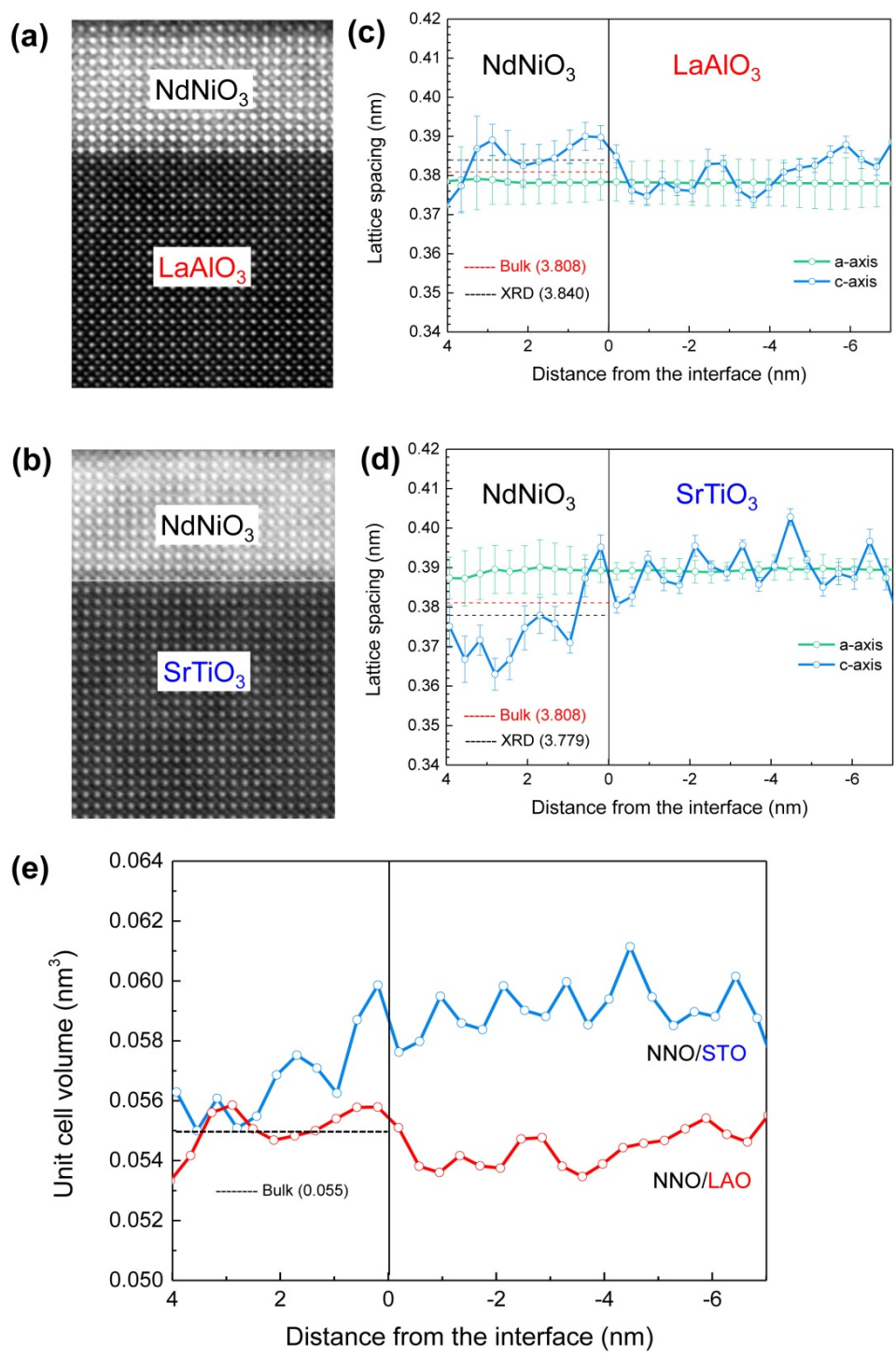

**Supplemental FIG. S6** (color online). STEM-high angle annular dark field (HAADF) images of the 10-u.c.-thick NdNiO$_3$ films grown on (a) LaAlO$_3$ and (b) SrTiO$_3$ substrates along the pseudocubic [100] zone axis. (c,d) The thickness profile of in-plane and out-of-plane lattice constants obtained across the NdNiO$_3$ films shown in (a) and (b), respectively. The red and black dashed lines represent the pseudocubic lattice constants of bulk NdNiO$_3$ and of as-grown

NdNiO$_3$ (001) films measured by XRD $\theta$-2$\theta$ scans, respectively. (e) The thickness-dependent unit-cell volumes of NdNiO$_3$/LaAlO$_3$ (red) and NdNiO$_3$/SrTiO$_3$ (blue) films derived from the lattice constants measured in (c) and (d). The black dashed line represents the pseudocubic unit-cell volume of bulk NdNiO$_3$. It is clear that the unit-cell volume in the NdNiO$_3$ film grown on a SrTiO$_3$ substrate expands compared with the bulk NdNiO$_3$, while that in the NdNiO$_3$ film grown on a LaAlO$_3$ substrate is close to the bulk value.

### RSMs of NdNiO$_3$ (001) films on YAlO$_3$ and SrTiO$_3$ substrates

To examine the effect of biaxial in-plane strain on the unit-cell volume of NdNiO$_3$ (001) films, we epitaxial grew two NdNiO$_3$/YAlO$_3$ and NdNiO$_3$/SrTiO$_3$ films, where the NdNiO3 films have lattice mismatches of -2.7 (in-plane compressive strain) and +2.5 % (in-plane tensile strain) with respect to the underlying YAlO$_3$ and SrTiO$_3$ substrates, respectively. By performing pseudocubic (-103) RSMs of these two NdNiO$_3$ films, we extracted the in-plane and out-of-plane lattice constants and calculated a unit-cell volume with the measured lattice constants. It is very interesting that the NdNiO$_3$/SrTiO$_3$ film is coherent with a volume expansion, whereas the NdNiO$_3$/YAlO$_3$ film is relaxed, preserving the unit-cell volume (Supplemental Fig. S7). In the NdNiO$_3$/SrTiO$_3$ film, oxygen vacancies are formed to accommodate the misfit strain (i.e. in-plane tensile strain) from the underlying SrTiO$_3$ substrate, resulting in the volume expansion of the NdNiO$_3$ film.

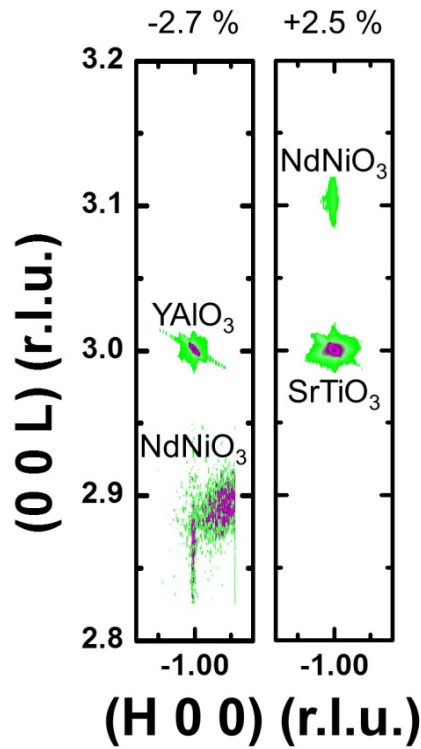

**Supplemental FIG. S7** (color online). (-103) RSMs of pseudocubic NdNiO$_3$ (001) films grown on YAlO$_3$ (left) and SrTiO$_3$ (right) substrates.

**Strain dependence of metal-to-insulator transitions (MITs) in NdNiO$_3$ (001) films**

To investigate the misfit-strain dependence of MITs in NdNiO$_3$ (001) films, we epitaxially grew various NdNiO$_3$ films with different strain states and then, measured the temperature ($T$)-dependent resistivity ($\rho$) of as-grown NdNiO$_3$ films in the *Van der Pauw* geometry. Note that NdNiO$_3$ films are under in-plane compressive (tensile) strain for YAlO$_3$ and LaAlO$_3$ (NdGaO$_3$, SrTiO$_3$, DyScO$_3$, and GdScO$_3$) substrates. By taking the derivative [$d(\log \rho)/dT$] of logarithmic resistivity (log $\rho$) with respect to $T$, we extracted MIT temperatures ($T_{MIT}$) for various NdNiO$_3$ films and plotted the $T_{MIT}$ as a function of in-plane biaxial strain. It was found that the MIT in the tensile-strained NdNiO$_3$ films occurs at a higher temperature, whereas it is suppressed in the compressive-strained NdNiO$_3$ films exhibiting a lower transition temperature (Supplemental Fig. S8). Intriguingly, for an NdNiO$_3$/YAlO$_3$ film, there is no MIT in the transport property, that is, the film was metallic at all temperatures.

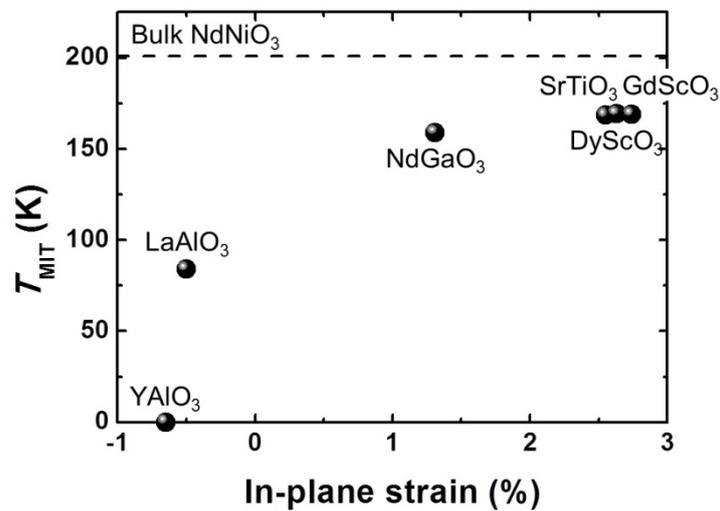

**Supplemental FIG. S8** (color online). The strain-dependent MIT temperatures $T_{MIT}$ in NdNiO$_3$ (001) films epitaxially grown on various substrates.

# Temperature-dependent Hall measurements of NdNiO$_3$/LaAlO$_3$ and NdNiO$_3$/SrTiO$_3$ films

To examine the effect of epitaxial strain on electronic properties in NdNiO$_3$ (001) films, we performed temperature-dependent Hall measurements of NdNiO$_3$/LaAlO$_3$ (in-plane compressive strain) and NdNiO$_3$/SrTiO$_3$ (in-plane tensile strain) films. From temperature-dependent plots of sheet resistance in these two NdNiO$_3$ films, we found that the NdNiO$_3$/SrTiO$_3$ film under in-plane tensile strain is electrically less conductive with larger sheet resistances at all temperatures than the NdNiO$_3$/LaAlO$_3$ film under in-plane compressive strain (Supplemental Fig. S9(a)). Through subsequent temperature-dependent Hall measurements, we also measured the concentration ($n$) of mobile charge carriers for both NdNiO$_3$/LaAlO$_3$ and NdNiO$_3$/SrTiO$_3$ films (Supplemental Fig. S9(b)). The NdNiO$_3$/LaAlO$_3$ film with better electrical conductivity exhibited higher carrier concentrations than the NdNiO$_3$/SrTiO$_3$ film at all temperatures. Strain-induced oxygen vacancies in the NdNiO$_3$/SrTiO$_3$ film may create insulating sites with band-gap opening locally, leading to the reduction of overall electrical conductivity.

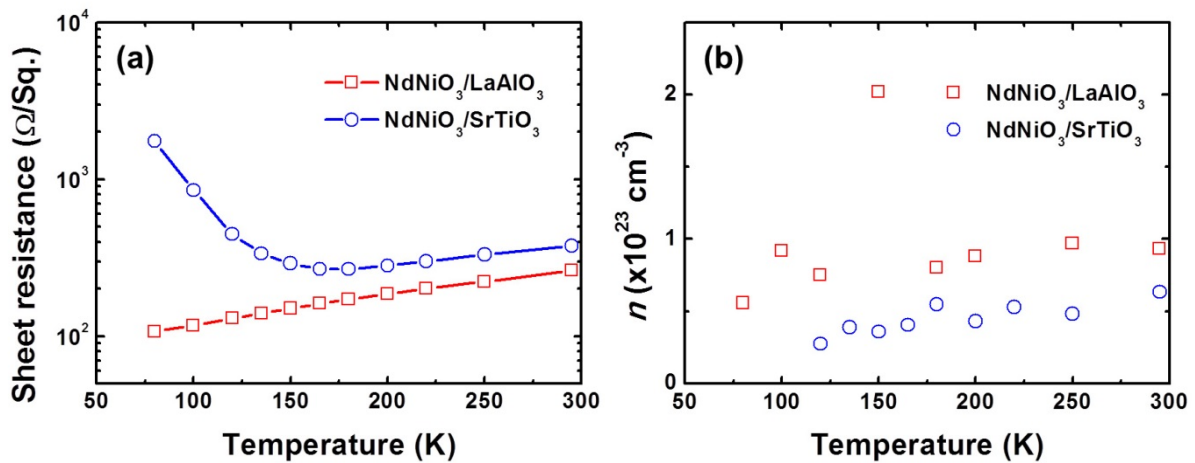

**Supplemental FIG. S9** (color online). The temperature-dependent (a) sheet resistance and (b) carrier concentration ($n$) in NdNiO$_3$ films epitaxially grown on LaAlO$_3$ (open red squares) and SrTiO$_3$ (open blue circles) (001) substrates.

**Cathodoluminescence (CL) measurements of as-grown NdNiO$_3$ (001) films**

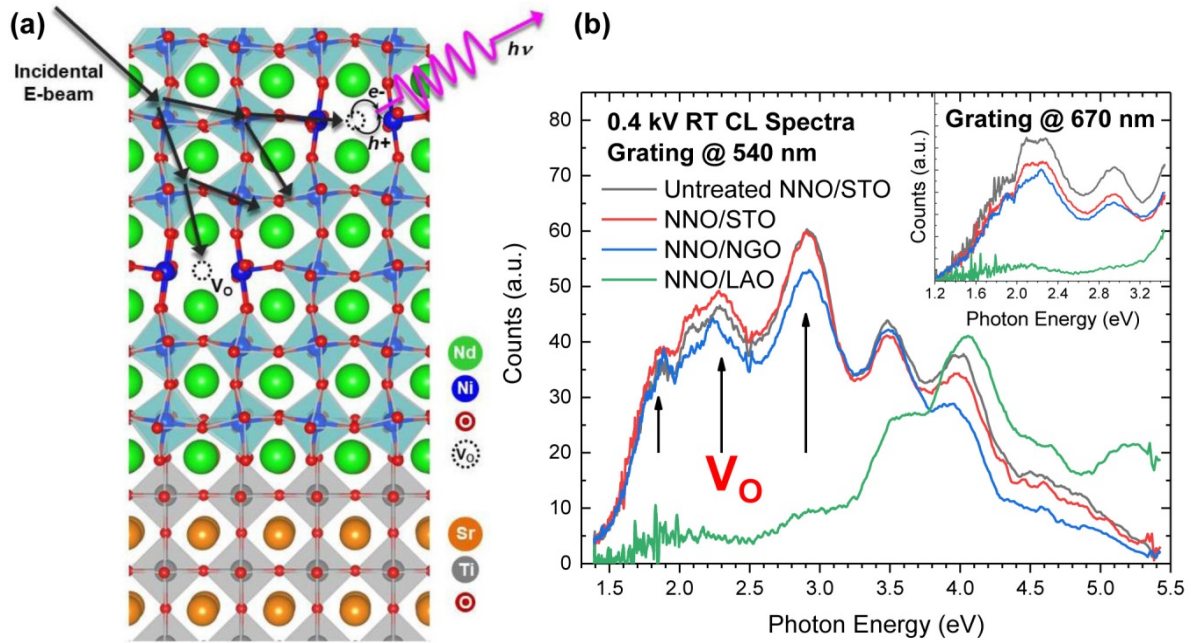

**Supplemental FIG. S10** (color online). (a) A schematic diagram of cathodoluminescence (CL) measurements in NdNiO$_3$ (001) films. (b) The CL spectra of NdNiO$_3$/LaAlO$_3$ (green), NdNiO$_3$/NdGaO$_3$ (blue), NdNiO$_3$/SrTiO$_3$ (red), NdNiO$_3$/SrTiO$_3$ (black) films *in situ* annealed under the reduced oxygen pressure (0.15 mbar). In the only CL spectra of tensile-strained NdNiO$_3$ films (i.e. NdNiO$_3$/NdGaO$_3$ and NdNiO$_3$/SrTiO$_3$ films), sizeable CL signals appear at the low photon energies of 1.8 eV which are directly related with oxygen-vacancy-mediated optical transitions.

**Atomic-level visualization of oxygen vacancies in NdNiO$_3$ (001) films under in-plane tensile strain**

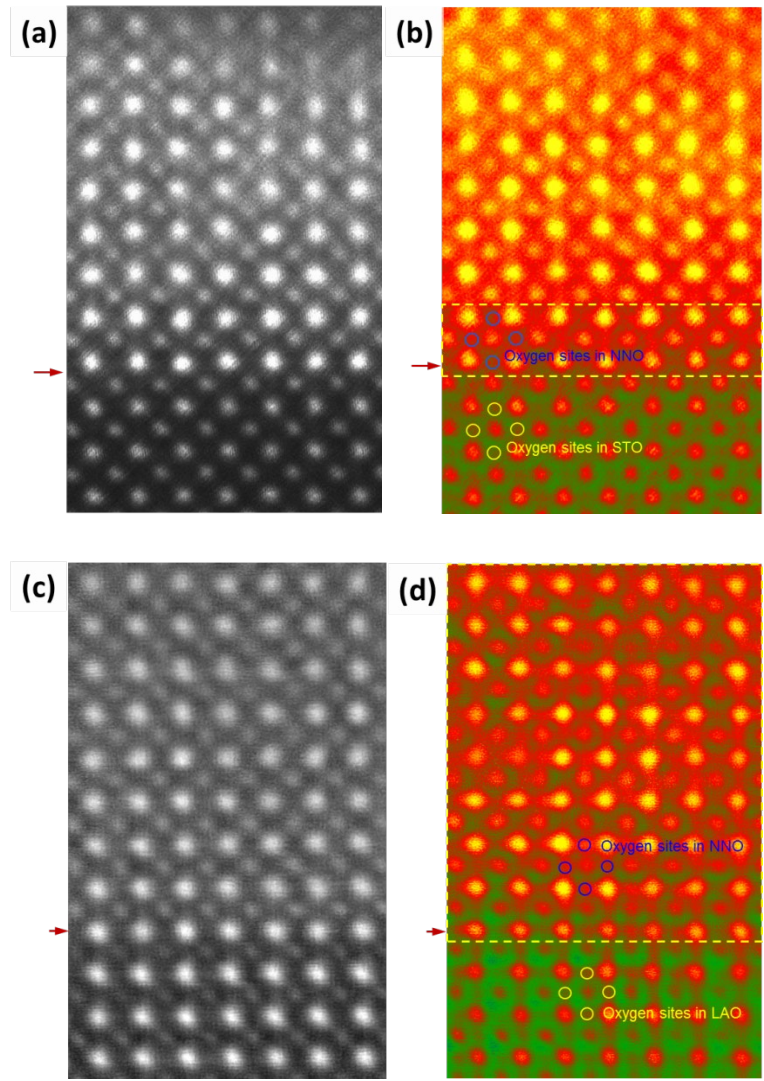

**Supplemental FIG. S11.** (color online). (a) and (c) high angle annular dark field (HAADF) and (b) and (d) annular bright field (ABF) STEM images of NdNiO$_3$ on SrTiO$_3$ and LaAlO$_3$ substrates, respectively. The arrows indicate the interface. While the oxygen atomic columns can be visualized as indicated by the yellow circles in SrTiO$_3$ substrate, LaAlO$_3$ substrate, and NdNiO$_3$ on LaAlO$_3$, the oxygen atoms hardly appear in the tensile-strained NdNiO$_3$ on SrTiO$_3$, which implies that the the oxygen atomic positions are deviated by the local disproportionation and therefore the dechanneling of the oxygen atoms prohibits the oxygen contrast in ABF STEM.

**Effect of oxygen vacancies on physical properties in NdNiO$_3$/SrTiO$_3$ (001) films**

To understand the role of oxygen vacancies in the physical properties of $R$NiO$_3$ films further, we prepared two NdNiO$_3$/SrTiO$_3$ (001) films with different oxygen stoichiometry (Supplemental Fig. S12(a)). After *in situ* film deposition, the oxygen contents of as-deposited NdNiO$_3$ films were manipulated by oxygen-ambient environments in a cooling process. An NdNiO$_3$ sample was cooled down under the oxygen partial pressure of 0.15 mbar and the other NdNiO$_3$ sample under the oxygen partial pressure of 1 atm. It should be noted that the NdNiO$_3$ film synthesized under the reduced oxygen pressure is more oxygen-deficient than that under the oxygen atmosphere pressure. And, both NdNiO$_3$ samples were structurally coherent with respect to the underlying SrTiO$_3$ substrates. But, a difference in the Bragg peak position is evident along the [00$L$] direction. The more oxygen-deficient an NdNiO$_3$ film is, the further a unit cell is elongated in the out-of-plane direction by oxygen vacancies. It is very interesting that the NdNiO$_3$/ SrTiO$_3$ film grown under the reduced oxygen pressure exhibits a larger out-of-plane lattice constant and thereby, a bigger unit-cell volume than the NdNiO$_3$/ SrTiO$_3$ film grown under the oxygen atmosphere pressure. Subsequent transport measurements also showed that the more oxygen-deficient an NdNiO$_3$ film is, the worse electrical conductivity is (Supplemental Fig. S12(b)). In bulk $R$NiO$_3$, a similar effect of oxygen vacancies on the physical properties was previously reported such as the expansion of a unit-cell volume and the degeneration of electrical conduction [Ref. 26 in the manuscript].

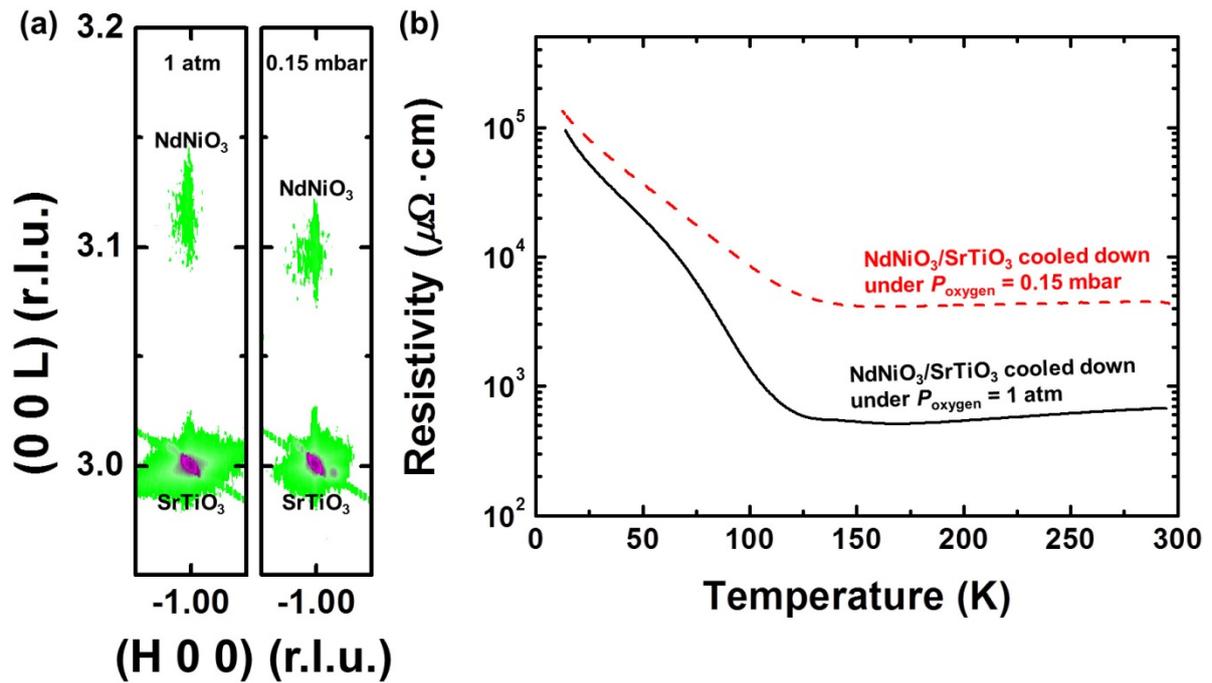

**Supplemental FIG. S12** (color online). (a) Pseudocubic (-103) RSMs of as-grown NdNiO$_3$/SrTiO$_3$ (001) films cooled down under the oxygen partial pressure ($P_{oxygen}$) of (left) 1 atm and (right) 0.15 mbar after *in situ* film deposition. (b) The temperature-depedent resistivity of the NdNiO$_3$/SrTiO$_3$ films cooled down under (the solid line) the oxygen atmosphere pressure (1 atm) and (the dashed line) the reduced oxygen pressure (0.15 mbar).

# Rutherford backscattering spectrometry (RBS) of NdNiO$_3$ (001) films on YAlO$_3$ and SrTiO$_3$ substrates

To exclude a possibility of cation vacancies in NdNiO$_3$ films, we carried out Rutherford backscattering spectrometry (RBS) measurements of epitaxial NdNiO$_3$/YAlO$_3$ (in-plane compressive strain) and NdNiO$_3$/SrTiO$_3$ (in-plane tensile strain) films (Supplemental Fig. S13). For these two as-grown NdNiO$_3$ films under the in-plane compressive and tensile strain, we found that the composition ratio between Nd and Ni atoms appeared to be 1 within the measurement errors of ± 0.5 and ± 1.0 %, respectively. Thus, we can conclude that the expansion of a unit-cell volume in NdNiO$_3$ films under in-plane tensile strain is mainly attributed to oxygen vacancies, not cation vacancies.

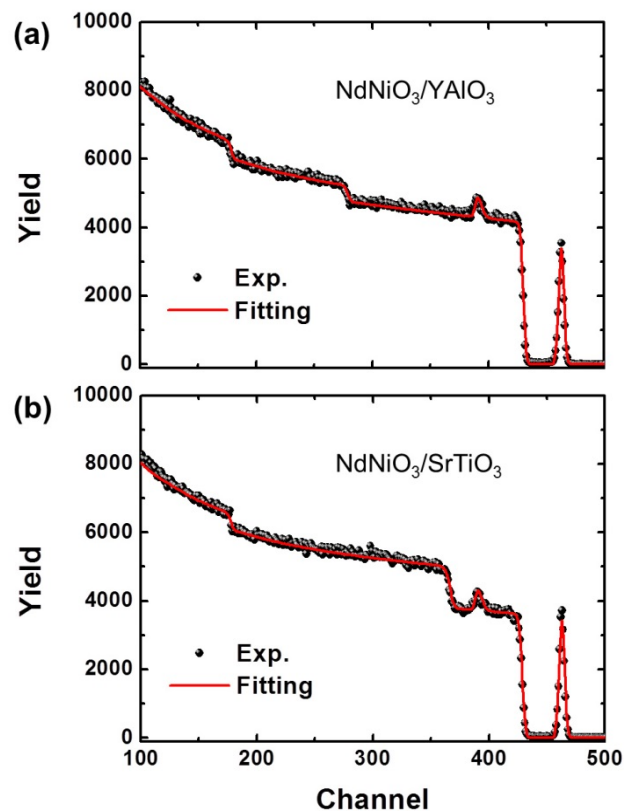

**Supplemental FIG. S13** (color online). RBS spectra of epitaxial (a) NdNiO$_3$/YAlO$_3$ and (b) NdNiO$_3$/SrTiO$_3$ films. The solid red lines show the best fit of the measured RBS spectra in (a) and (b).

# X-ray absorption spectroscopy (XAS) of 14-nm-thick NdNiO$_3$ (001) films

X-ray absorption spectra (XAS) were measured at the REIXS beamline of the Canadian Light Source. Spectra were measured using bulk-sensitive, total fluorescence yield (TFY) detection, using a micro-channel plate detector, with an energy resolution of approximately 0.1 eV. During measurements, the incident photon beam was 30 degrees from the sample surface, and reported spectra are an average over linear vertical and linear horizontal polarization.

Supplemental Figure S14(a) shows Ni $L_{2,3}$ x-ray absorption spectroscopy (XAS) data for the 14-nm-thick NdNiO$_3$ films on different substrates. For each case, spectra measured above and below the MIT temperature show pronounced changes typical of nickelates [Ref. 42,43 in the manuscript]. In the insulating phase, a two-peaked structure is present at the Ni $L_3$ edge (~853 eV). However, we note the first sharp peak (denoted by the dotted line in Supplemental Fig. S14(a)) becomes stronger and shifts to lower energy as the substrate changes from LaAlO$_3$ (LAO) to NdGaO$_3$ (NGO) to SrTiO$_3$ (STO). While the sharp peak is an intrinsic feature of the nickelate spectrum [Ref. 6,42–44 in the manuscript], the similarly sharp $L_3$ peak of Ni$^{2+}$ in NiO is only slightly lower in energy. Thus, the increase of peak intensity and shifting to lower energy with tensile strain indicates a reduction of the Ni valence in conjunction with the introduction of oxygen vacancies.

In the anti-ferromagnetic ordered phase, the (1/4,1/4,1/4) magnetic Bragg peak can be observed when using photons tuned in energy to the Ni $L_3$ edge, and the energy dependence of this peak is dependent on the degree of disproportionation [Ref. 6 in the manuscript]. Supplemental Figure S14(b) shows energy scans of this resonant magnetic diffraction (RMD) intensity across the Ni $L_3$ edge for the three films, collected using a photo-diode positioned according to the magnetic Bragg condition. In accordance with the growth and shift of the first XAS peak to lower energies for tensile strain, we observe that the RMD peak also shifts to lower energies. The RMD probes the long range magnetic ordering, and thus this observation is a verification of the enhanced disproportionation over long ranges predicted by the DFT calculations [Ref. 6 in the manuscript]. Lastly, we show oxygen $K$ edge XAS for the three films, where again a shift to lower energies is observed for the films on NGO and STO substrates (Supplemental Fig. S14(c)). This shift to lower energy for the oxygen $K$ edge XAS of tensile strained films is in agreement with previous studies on NdNiO$_3$ [Ref. 17 in the manuscript].

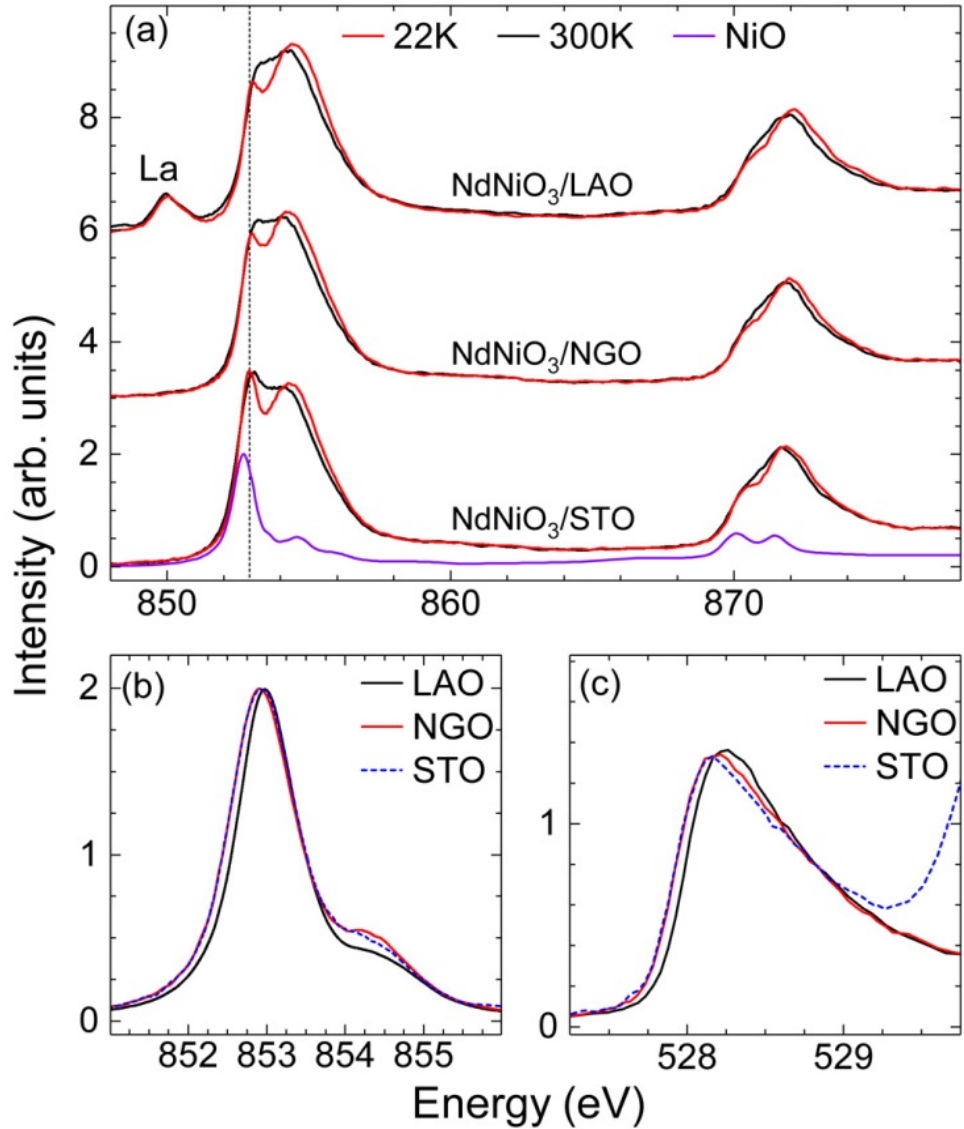

**Supplemental FIG. S14** (color online). (a) Ni $L_{2,3}$ x-ray absorption spectroscopy (XAS) of 14-nm-thick NdNiO$_3$ films on LAO, NGO, and STO substrates below (22 K) and above (300 K) the MIT. (b) Energy dependence of the resonant magnetic scattering for the three films at the Ni $L_3$ edge at 22 K. (c) Oxygen $K$ edge XAS at 22 K.

## The effect of oxygen vacancies on the Ni charge valency in NdNiO$_3$ (001) films

To check the effect of oxygen vacancies on the oxidation states of Ni ions, we performed Ni $L_{2,3}$ x-ray absorption spectroscopy (XAS) measurements of two NdNiO$_3$/SrTiO$_3$ films with different oxygen stoichiometry (Supplemental Fig. S15). Here, the one (the blue curve, less oxygen-deficient) was synthesized with the *in situ* post-annealing of 1 atm after film deposition under the oxygen partial pressure of 0.15 mbar. The other (the magenta curve, more oxygen-deficient) was cooled down without any oxygen post-annealing for the as-deposited film. It is evident that the more oxygen-deficient the as-grown film is, the closer its XAS spectra is to that of NiO. This implies that oxygen vacancies in NdNiO$_3$ films are easily formed with in-plane tensile strain, which leads to the reduction of the Ni charge valence from covalent Ni$^{3+}$ to ionic Ni$^{2+}$.

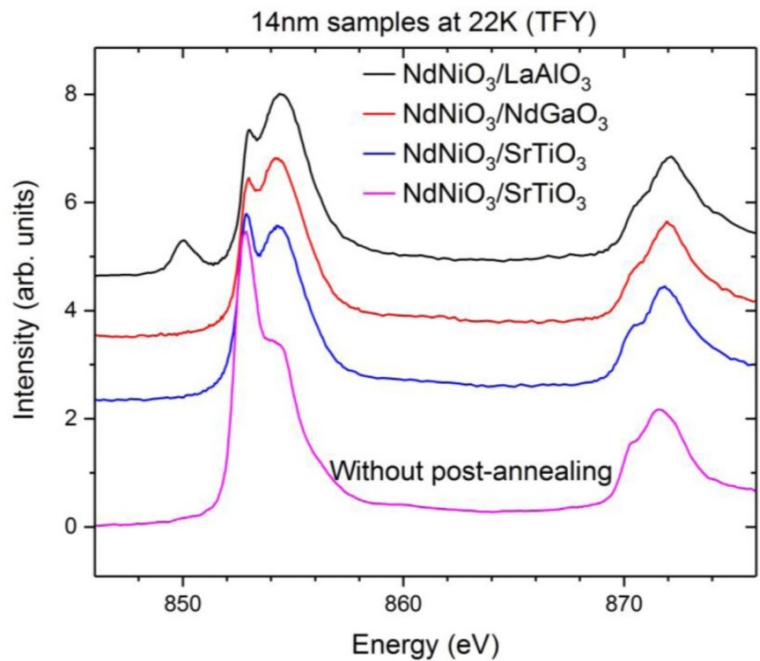

**Supplemental FIG. S15** (color online). Ni $L_{2,3}$ X-ray absorption spectroscopy (XAS) of 14-nm-thick NdNiO$_3$ films on LaAlO$_3$ (black), NdGaO$_3$ (red), and SrTiO$_3$ (blue) substrates at 22 K. The NdNiO$_3$ films are *in situ* post-annealed under an oxygen partial pressure of 1 atom. In contrast, an oxygen-deficient NdNiO$_3$/SrTiO$_3$ (magenta) film is intentionally prepared by cooling down the as-deposited film without *in situ* oxygen post-annealing.